# Decoding MGMT Methylation: A Step Towards Precision Medicine in Glioblastoma

Hafeez Ur Rehman[∥,†,‡], Sumaiya Fazal[†,¶], Moutaz Alazab[*,‡,§], Ali Baydoun[‡]

[†]Department of Computer Science, National University of Computer and Emerging Sciences, Islamabad, Pakistan.
[‡]School of Computing and Data Sciences, Oryx Universal College with Liverpool John Moores University, Doha, Qatar.
[§] Department of Intelligent Systems, Faculty of Artificial Intelligence, Al-Balqa Applied University, Al-Salt 19117, Jordan.
[*] Corresponding Author; Email: m.alazab@bau.edu.jo,
[∥] hafeez.urrehman@nu.edu.pk, hafeez.r@oryx.edu.qa, h.urrehman@ljmu.ac.uk
[¶] p200401@nu.edu.pk

*Abstract*—Glioblastomas, constituting over 50% of malignant brain tumors, are highly aggressive brain tumors that pose substantial treatment challenges due to their rapid progression and resistance to standard therapies. The methylation status of the O-6-Methylguanine-DNA Methyltransferase (MGMT) gene is a critical biomarker for predicting patient response to treatment, particularly with the alkylating agent temozolomide. However, accurately predicting MGMT methylation status using non-invasive imaging techniques remains challenging due to the complex and heterogeneous nature of glioblastomas, that includes, uneven contrast, variability within lesions, and irregular enhancement patterns. This study introduces the Convolutional Autoencoders for MGMT Methylation Status Prediction (CAMP) framework, which is based on adaptive sparse penalties to enhance predictive accuracy. The CAMP framework operates in two phases: first, generating synthetic MRI slices through a tailored autoencoder that effectively captures and preserves intricate tissue and tumor structures across different MRI modalities; second, predicting MGMT methylation status using a convolutional neural network enhanced by adaptive sparse penalties. The adaptive sparse penalty dynamically adjusts to variations in the data, such as contrast differences and tumor locations in MR images. Our method excels in MRI image synthesis, preserving brain tissue, fat, and individual tumor structures across all MRI modalities. Validated on benchmark datasets, CAMP achieved an accuracy of 0.97, specificity of 0.98, and sensitivity of 0.97, significantly outperforming existing methods. These results demonstrate the potential of the CAMP framework to improve the interpretation of MRI data and contribute to more personalized treatment strategies for glioblastoma patients.

## I. Introduction

A brain tumor is an abnormal accumulation of brain cells that can be either malignant or benign. While benign tumors are not carcinogenic, malignant tumors can be. Primary and secondary kinds of brain tumors are also distinguished. Primary brain tumors are often benign, however, subsequent brain tumors are typically malignant [1]. The World Health Organization (WHO) divides brain tumors into four categories, ranging in severity from grade I to IV. But some brain tumors, such as gliomas, are more frequent [1]. The most frequent primary tumor is a glioma, which can develop in the brain or spine. Depending on how malignant they are, gliomas have been further categorized by the WHO into several classes.

The grading system for brain tumors spans from Grade I, the least aggressive grade, through grade IV, the most aggressive grade. Grade I denotes a primary brain tumor, while grades II and III are classified as malignant, with grade IV also known as glioblastoma multiforme (GBM)—being associated with the highest risk.

The most typical and invasive type of malignant brain tumor is glioblastoma, a grade IV brain tumor. The majority of glioblastoma cases involving the central nervous system (CNS) in adults are caused by the primary subtype, according to [2]. The shape, appearance, and histology of glioblastomas are all naturally variable. Patients with glioblastoma often have an average prognosis of 14 months with standard treatment, including chemotherapy and radiotherapy, due to their extremely infiltrative nature. The heterogeneity of the tumor makes it difficult to diagnose, prognosis, and forecast survival solely on MRI images.

The necessity for a change in the clinical diagnosis of malignancies was stressed in a recent update to the World Health Organization (WHO) classification of central nervous system (CNS) tumors. This change implies using molecular and cytogenetic features rather than only depending on morphological and histological traits for classification and prognosis [3]. The presence of O6-Methylguanine-DNA Methyltransferase (MGMT) in glioblastoma, an aggressive kind of brain cancer, is crucial in evaluating the efficacy of treatment. As a "suicide" DNA repair enzyme, MGMT does its job. Methylation of the MGMT promoter region improves therapeutic efficacy by preventing the DNA sequence's O6 guanine from being broken by the temozolomide alkylating agent. A positive prognostic indicator for the effectiveness of treatment is the MGMT status.

Surgery or biopsy are frequently used to determine the status of O6-Methylguanine-DNA Methyltransferase (MGMT). However, establishing an accurate genetic characterization using these techniques can be time-consuming, necessitating additional surgical procedures and many weeks for analysis.

Furthermore, because of the heterogeneity of glioblastoma (GBM), biopsy samples are subject to variation across different evaluators and within a single evaluator, which can result in a downgrade in tumor rating, as indicated in prior study [4].

In order to overcome these difficulties, a non-invasive method of MGMT status determination without the need for biopsy techniques is required. In comparison to other imaging modalities, magnetic resonance imaging (MRI) provides better image contrast for seeing brain tissues [5], [6]. Modern machine learning methods combined with MRI images have the potential to significantly reduce the need for surgical resection and enable accurate prediction of MGMT methylation status.

Numerous studies have proposed various methods for predicting the prognosis of brain cancers based on MGMT status, demonstrating significant discrepancies [4], [7]. Ellingson et al. [8] have proposed that detectable genetic abnormalities within glioblastomas are macroscopic and can be found using MRI in the field of radiogenomics. Moon et al. [9] have noticed that, on the other hand, MGMT methylation frequently correlates with diffuse tumor boundaries.

In the first radiogenomics investigation, MRI characteristics and gene expression were connected to find novel molecular determinants. Gautam et al. used quantitative characteristics generated from sub-region segmentation, such as edema, necrosis, and tumor-enhancing regions [10], [11]. This study found important connections between molecular profiles. The selection of quantitative factors, however, presents a significant issue because it calls for the experience of seasoned radiologists for direction.

Numerous studies [7], [9] have demonstrated a relationship between MGMT methylation status and MR images. According to their findings, the methylation state of MGMT is correlated with particular tumor sites [7]. Additionally, researchers have connected computed tomography data to MGMT methylation status. The association with MGMT methylation status has been investigated using subsequent MRI sequences, such as T2, Flair, post-contrast T1 (Gd), MPMRI, post-contrast T1 weighted (T1 Gd), T2 weighted, and T2 FLAIR.

By utilizing imaging data, many researchers have started their investigations into genetic analysis, notably the evaluation of MGMT promoter methylation status. The availability of current datasets that connect diverse genetic markers with related photos has sparked this movement. Magnetic resonance imaging (MRI) stands out among the wide range of image formats as being particularly skilled in revealing MGMT promoter methylation status. The prediction of MGMT promoter methylation from MRI data has been the subject of numerous investigations, with writers using spatial frequency texture analysis to produce their conclusions [12], [13]. To capture fluctuations in image intensities within magnetic resonance (MR) images of patients suffering from glioblastoma multiforme (GBM), they have specifically used run length matrices based on texture features [14].

One significant example is the correlation between computed tomography (CT) image characteristics and MGMT promoter methylation [7]. It is important to highlight that, because to the greater soft tissue contrast offered by MR images, CT-derived variables and those computed from MR images cannot be directly compared. Although it can be difficult to manually detect clinical characteristics from MRI scans, empirical data from earlier studies has shown that MR images have the ability to manifest apparent distinctions in predicting MGMT methylation status [4], [7].

For computational methods that rely on MR images, glioblastoma (GBM) heterogeneity refers to the tumor's heterogeneous composition and features. These computational methods have trouble capturing the complex variations that exist. A number of researchers worked together to create an automated Glioblastoma survival prediction model in an effort to address this problem. Their efforts produced promising outcomes that showed a fair amount of accuracy. They were able to do this by combining different textures and features, using both a cutting-edge neural network architecture and a random forest regression model [15].

Deep learning approaches have been used more frequently recently for glioblastoma status prediction and localization, with encouraging results [16], [17]. Positive results, especially on a small validation dataset [18], were obtained by using a bi-directional convolutional recurrent neural network technique that takes into consideration the spatial properties of 3-dimensional MRI images.

Three different residual deep neural networks (ResNet) were used by a group of researchers in a ground-breaking study to evaluate how well each network performed in predicting the MGMT status [19]. Notably, this method avoided the need for additional pre-processing steps by skipping the segmentation stage. The ResNet50 architecture had the most outstanding accuracy out of the three options. However, it's important to remember that the dataset for this study was limited to a few people. An investigation of the use of autoencoders in the field of medical imaging was recently conducted by [20]. Autoencoders are a popular tool in computer science research because of their exceptional capacity for unsupervised learning to provide useful data representations. These neural network topologies are essential for data compression, feature extraction, and dimensionality reduction. Autoencoders have the potential to be a useful tool for understanding the properties of MR images and predicting the relationship between image qualities and the MGMT methylation status since they can detect hidden patterns and attributes within data.

Intricate patterns in genomic or epigenomic data are expertly recognized by autoencoders, making it easier to extract important aspects that might otherwise be challenging to identify. By utilizing autoencoders, it may be possible to identify tiny chemical fingerprints related to the methylation status of the MGMT promoter, greatly improving prediction accuracy.

The inconsistent and varied appearance of methylation signals, which current computational approaches struggle to resolve, poses a substantial barrier in determining MGMT status using MR images. It is already a difficult effort to predict the MGMT promoter methylation status, and it becomes significantly more difficult when dealing with several tumor locations, significant heterogeneity within lesions, and various enhancing patterns . In order to overcome these difficulties, we present Convolutional Autoencoders for Methylation Status Prediction (CAMP) using adaptive penalty, a unique technique that analyzes the signatures found in both normal and glioblastoma-containing MRI images to estimate the methylation status of the MGMT promoter gene.

Our strategy initially includes repeatedly creating synthetic MRI image using convolutional autoencoder to identify hidden image characteristics as It effectively preserves brain tissue, fat, and individual tumor structures across various MRI modalities. This enhancement is crucial for accurate medical analysis and treatment planning. Secondly, in order to predict the MGMT methylation status, The methodology incorporates an adaptive sparse penalty that dynamically adjusts to variations in features transferred from the latent space used to generate synthetic MRI, such as contrast differences and tumor locations in MR images. This feature enhances the model's ability to predict MGMT methylation status accurately. It is noteworthy that by using a convolutional autoencoder we found the most optimal results when predicting the MGMT methylation status. The framework achieves excellent performance metrics, with an accuracy score of 0.97, specificity of 0.98, and sensitivity of 0.97. These results indicate a high level of precision in predicting the MGMT methylation status, which is vital for effective glioblastoma treatment.

The manuscript is structured into five main sections. To begin, the first section provides an in-depth introduction to the problem at hand, supplemented by an extensive review of the relevant literature. Section 2 delineates the methodology we propose, while Section 3 offers an exposition of our discoveries and significant insights. Furthermore, the Discussion section elucidates how our proposed technique tackles the issues we have identified. Ultimately, the concluding section encapsulates our overall findings and conclusions drawn from this study.

## II. METHODOLOGY

The Convolutional Autoencoders for Methylation Status Prediction (CAMP) framework operates in two distinct phases. In the initial phase is CAMP-I where the primary focus is on enhancing the representation of MRI images and generating an innovative set of synthetic MR images. This process is instrumental in enabling the model to extract pertinent features relating to diverse tumor locations and intricate MGMT patterns effectively. Following the completion of synthetic image generation, the second phase convolutional neural network is used to make predictions regarding the MGMT status.

The CAMP framework incorporates a customized convolutional autoencoder. The subsequent subsections provide a more comprehensive breakdown of each of these phases.

The block diagram illustrating the initial phase (CAMP-I) of the proposed CAMP framework is presented in Figure 2. In CAMP-I, each input MR sequence is derived from a pool of 586 patients. This dataset is derived from MRI images of glioblastoma tumors, sourced from a publicly accessible Radiologist Society of North America (RSNA) database that incorporates MRI images coupled with gene expression status [21]. The MR sequences for each patient encompass fluid-attenuated-inversion recovery (FLAIR), T1-weighted (T1w), T1-weighted-contrast-enhanced (T1wCE), and T2-weighted (T2w) slices. Patients are categorized based on their MGMT methylation status. All MRI images have undergone independent evaluation by four certified neuroradiologists for each patient, with the neuroradiologists being blind to the MGMT status of the tumors. In total, the dataset comprises 586 patients. Each individual sample comprises 512x512 intensity values, ranging from 0 to 255. Many of these samples contain minimal or no relevant information. Therefore, as an initial preprocessing step, we perform resizing to 256x256 and slice selection to extract slices containing substantial information.

*1) MRI Slice Pre-Processing:* To extract slices brimming with meaningful information, a sophisticated slice selection procedure is executed. This pivotal step hinges upon the informational depth of each slice, determining whether to retain or discard it. Understanding the limitations of solely relying on a singular metric, particularly in scenarios tainted by noisy slices such as those distorted by Gaussian noise, a two-pronged metric strategy is championed. This involves harnessing the strengths of both entropy (H) and signal-to-noise ratio (SNR). Implementing this dual-metric approach guarantees the retention of slices that are both informative and primed for the next stages of processing.

Entropy (1) serves as a lens to evaluate the unpredictability or randomness present within an MR slice. To ascertain the entropy value of a given slice $x$, symbolized as $H(x)$, the subsequent equation 1 is applied:

$$H(x) = -\sum_{j=0}^{D-1} p_j \log_2(p_j(x)) \quad (1)$$

In this context, $D$ symbolizes the comprehensive number of unique intensity levels within the slice, which, for our purposes is fixed at 256. The term $j$ denotes individual shades, and $p_j$ is the likelihood of the shade $j$ materializing within the slice $x$. Within our dataset, the entropy values for MR slices span between 0 to 3.2, indicating that not all slices explore the full-intensity scope.

Post-experimental analysis on an assortment of patient slices revealed an optimal entropy benchmark of 1.3. From both a visual standpoint and for model optimization, slices surpassing this entropy threshold are designated. Bypassing MRI scans beneath this marker not only elevates the caliber of slices chosen but also judiciously economizes training resources.

Our secondary yardstick, the signal-to-noise ratio (SNR) (2), assists in pinpointing information-dense slices. For an input slice, signified as $x$, the SNR is deduced via:

$$\text{SNR}(x) = 10\log_{10}\frac{\bar{e}(x)}{\sigma^2(x)} \quad (2)$$

In equation 2, $\bar{e}(x)$ represents the average luminosity of an MR slice, derived from the squared mean pixel value. In contrast, noise is delineated by the pixel value fluctuations within a slice, represented by $\sigma^2(x)$. For an accurate noise evaluation, one should focus on a consistent gray-colored, sufficiently expansive region within the image. This particular region, typically a flat border segment, is leveraged to calculate noise parameters.

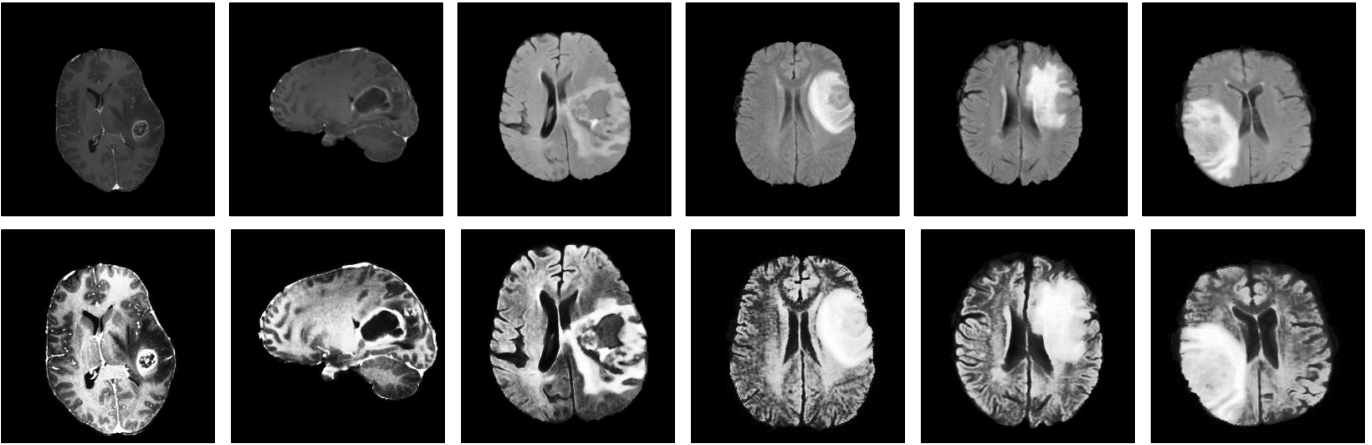

Fig. 1: (a) Row 1: Original Images of Various Modalities(T1w, T2w, FLAIR, T1wCE). (b) Row 2: The corresponding preprocessed images of various modalities(T1w, T2w, FLAIR, T1wCE) revealing significant details even of low contrast regions.

Slices that fail to meet the SNR threshold, labeled 'T', are eliminated from the selection pool. The threshold 'T' is meticulously chosen after rigorous experimentation on a subset of random MRI slices.

However, the chosen MR slices do display an array of non-uniform intensity variations. Some appear faint, others muted, and a significant fraction lacks sharp contrast, as shown in Figure 1. To counteract these discrepancies, we employ the histogram equalization method. This technique amplifies the contrast by realigning the intensity levels, aiming for a more balanced distribution across the spectrum. This modification ensures slices bear a more consistent intensity profile, resulting in sharper contrast and clarity. The transformative effect of this method on several slices is evident in the succeeding section of Fig. 1.

*CAMP-I: Feature Learning*

The image slices that we've improved from the previous phase of preprocessing are now set as the foundation for our specially designed convolutional autoencoder. Autoencoders are like multitool of machine learning, known for their wide range of applications. Especially in situations where you don't have labeled data, autoencoders shine by learning about the data without needing examples [20]. In short, they're great at picking out the most important features from data, which can later be used for various purposes.

Convolutionals neural networks (CNNs) are used as stacks of layers, each designed to do various tasks. Some layers learn features, others reduce data size, and some can help in decision-making. CNNs stand out, especially for images, in how they use special operations to focus on small patterns and then combine them to understand bigger patterns. Pairing these CNN layers with autoencoders is an advanced tool that can automatically figure out the important parts of images. This makes our encoder really efficient at taking an image, and shrinking it down while keeping all the important details.

In order to capture intricate details of MGMT patterns in MR images, we employed Convolutional Autoencoders tailored for data compression through an encoder and subsequent reconstruction via a decoder. A customized convolutional autoencoder design, detailed in Table I, was developed to address the specific requirements of the dataset. The primary goal was to minimize discrepancies between the input and output, ensuring the reconstructed output closely resembled the input. This approach aimed to enhance the system's ability to comprehend the subtle details present in MR slices.

TABLE I: CAMP-I: Model Architecture for Feature Learning

| Layer (type) | Output Shape | Param # |
|---|---|---|
| InputLayer | [(None, 256, 256, 1)] | 0 |
| Conv2D | [(None, 256, 256, 64)] | 640 |
| MaxPooling2D | [(None, 128, 128, 64)] | 0 |
| Conv2D | [(None, 128, 128, 32)] | 18464 |
| MaxPooling2D | [(None, 64, 64, 32)] | 0 |
| Conv2DTranspose | [(None, 128, 128, 32)] | 9248 |
| Conv2DTranspose | [(None, 256, 256, 64)] | 18496 |
| Conv2D | [(None, 256, 256, 1)] | 577 |

Enhanced image slice is fed to the CNN, say about the size of 256×256 pixels. Every pixel in this image goes into our encoder. This encoder has layers that use special functions (LeakyReLU, in our case) to process the data. The formula, which you can see in equation 3:

$$y = \Re(\overline{w} \cdot \overline{X} + b) \quad (3)$$

describes this process. As we go deeper into the encoder, the data size gets smaller. This forces our system to focus on the most important details while leaving out the unnecessary ones. Fewer details mean the system learns faster and better. Plus, it's less work for our computer, making things efficient.

At the end of the encoder, we get a compact version of our image, which then enters the decoder. Here, the process is somewhat reversed, as the decoder tries to expand this small data back to its original form.

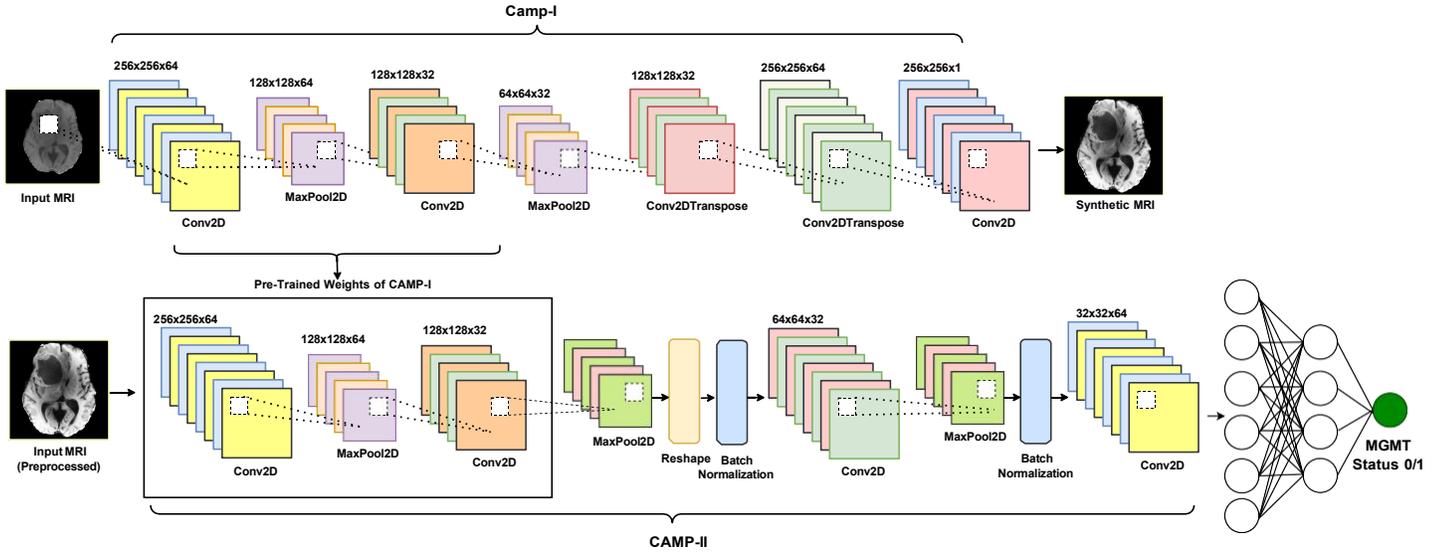

Fig. 2: A novel CAMP (Convolutional Autoencoder for MGMT Prediction) model designed for predicting MGMT methylation status in glioblastoma. The model leverages an Adaptive Convolutional Autoencoder architecture, showcasing its potential to efficiently capture intricate patterns and features essential for accurate prediction of glioblastoma using MR scans.

To make our system even smarter, we added a bit of noise to the input images. This makes our system focus more on the actual structure of the image rather than the noise. By doing this, our model gets really good at spotting the essential features of the image.

When we wanted to create new images, we used a metric called Dice loss, often seen in tasks where you need to segment or divide up images. Dice loss checks how close our newly created image is to a target image. You can see how this works in equation 4:

$$L_{Dice} = \frac{2 \sum_{i=1}^{N} G_i \cdot T_i}{\sum_{i=1}^{N} G_i^2 + \sum_{i=1}^{N} T_i^2} \quad (4)$$

In simpler terms, this equation checks how much overlap there is between the generated and target image pixels. So, with every pixel in the generated MR slice having a potential 256 shades, this equation helps ensure our created images are as accurate as possible.

*CAMP-II: Adaptive Penalty for Prediction of MGMT*

The second phase of the CAMP framework involves a convolutional neural network (CNN). The network takes as input an enhanced slice with dimensions of 256x256 pixels. Building upon the features learned in the first phase (CAMP-I), where artificial MRI images were generated iteratively, the second phase employs a customized convolutional neural network. This CNN aims to predict the MGMT methylation status from the signatures detected in MRI images.

The network excels at extracting informative features by initially selecting low-level features and then concurrently constructing features relevant to the original input. To ensure the selection of the most pertinent and optimized set of features. Additionally, transfer learning is utilized to predict the methylation status of the MRI slice. By regularizing the autoencoder network with sparsity constraints, it acquires the ability to learn unique statistical features from the dataset on which it has been trained. This goes beyond mere input replication. Training the autoencoder with a sparse penalty encourages it to uncover valuable features, thus enabling it to predict the MGMT methylation status more effectively.

The overall architecture of the proposed convolutional autoencoder network is illustrated in Figure 2. This architecture builds upon the foundation established in CAMP-I, where important features were learned and subsequently transferred to CAMP-II for MGMT methylation status classification following extensive training.

TABLE II: CAMP-II Model Architecture for MGMT Prediction

| Layer (type) | Output Shape | Param # |
| --- | --- | --- |
| InputLayer | (None, 256, 256, 1) | 0 |
| Conv2D | (None, 256, 256, 64) | 640 |
| MaxPooling2D | (None, 128, 128, 64) | 0 |
| Conv2D | (None, 128, 128, 32) | 18,464 |
| MaxPooling2D | (None, 64, 64, 32) | 0 |
| Reshape | (None, 64, 64, 32) | 0 |
| BatchNormalization | (None, 64, 64, 32) | 128 |
| Conv2D | (None, 64, 64, 32) | 4,128 |
| MaxPooling2D | (None, 32, 32, 32) | 0 |
| BatchNormalization | (None, 32, 32, 32) | 128 |
| Conv2D | (None, 32, 32, 64) | 8,256 |
| Flatten | (None, 65536) | 0 |
| Dense | (None, 64) | 4,194,368 |
| Dense | (None, 1) | 65 |

*2) Adaptive Sparse Penalty:* The Adaptive Sparse Regularizer (ASR) is a mathematically sophisticated approach designed to ensure optimal network activations by dynamically modulating their sparsity levels. Central to this regularization is the KL Divergence, which quantifies the difference between two probability distributions. In this context, the KL Divergence captures the gap between the desired sparsity level, denoted as $p$, and the actual sparsity, represented by $\hat{p}$. The equation for the divergence, given by:

$$KL(p||\hat{p}) = p \log \frac{p}{\hat{p}} + (1-p) \log \frac{1-p}{1-\hat{p}}$$

This divergence measure acts as a sentinel, monitoring the deviation of $\hat{p}$ i.e., the mean activation over a batch, to $p$ which is the the target activation threshold. The mean, i.e., $\hat{p}$ is the aggregated mean derived from the entire activation matrix.

Building upon this divergence, the adaptive nature of the regularizer emerges when determining the regularization strength, denoted as $\beta_2$. Instead of a fixed value, $\beta_2$ is determined dynamically, depending on the divergence:

$$\beta_2 = \beta_{min} + (\beta_{max} - \beta_{min}) \times KL(p||\hat{p})$$

.

Here, $\beta_{min}$ and $\beta_{max}$ represent the bounds within which the regularization strength oscillates. This adaptive determination of $\beta_2$ ensures that as the deviation between desired and actual sparsity grows, the penalty becomes more stringent, driving the network to adjust its activations accordingly.

The culmination of this intricate process is represented by the final regularization term, $R$. It's a composite metric, determined by multiplying the dynamically adjusted $\beta_2$ with the aggregate of KL divergences for all activation units:

$$R = \beta_2 \times \sum_{i=1} KL(p||\hat{p}_i)$$

.

In essence, the Adaptive Sparse Regularizer (ASR) leverages the principles of KL Divergence and dynamic regularization strength to ensure that the neural network maintains an optimal balance in its activations, avoiding both overfitting and underfitting. By dynamically penalizing deviations from the desired sparsity, it guides the network to focus on the most pertinent features, thereby enhancing its performance and ensuring better results.

Lastly, for an adaptive sparse autoencoder's loss function that comprises a reconstruction error, let's denote the reconstruction error as $L_{recon}$

$$L = L_{recon} + R$$

Where $R$ is the adaptive sparse regularizer as described above. This combined loss ensures not only that the network reconstructs the input data but also maintains the desired sparsity in its activations, enhancing the model's efficiency and generalization capabilities.

**Algorithm 1** Adaptive Sparse Regularizer
---
**Require:** *activation_matrix*
**Ensure:** $p \leftarrow 0.2$ {Target sparsity level}
**Ensure:** $\beta_{min} \leftarrow 1.00$ {Minimum regularization strength}
**Ensure:** $\beta_{max} \leftarrow 5.00$ {Maximum regularization strength}
1: $\hat{p} \leftarrow$ mean(*activation_matrix*)
2: local $\leftarrow$ std(*activation_matrix*)
3: n $\leftarrow$ getActivationCount()
4: *sum_KL* $\leftarrow 0$
5: *result* $\leftarrow 0$
6: **for** $i \leftarrow 1$ to $n$ **do**
7: $\quad$ KL divergence($p, \hat{p}_i$) $\leftarrow p * \log \frac{p}{\hat{p}_i} + (1-p) * \log \frac{1-p}{1-\hat{p}_i}$
8: $\quad$ *sum_KL* $\leftarrow$ *sum_KL* + KL_divergence($p, \hat{p}_i$)
9: **end for**
10: $\beta_2 \leftarrow \beta_{min} + (\beta_{max} - \beta_{min}) * sum\_KL$
11: *result* $\leftarrow \beta_2 * sum\_KL$
$\quad$ {Compute the regularization term}
12: **return** *result*

## III. EXPERIMENTAL SETUP

Our proposed CAMP framework operates in two distinct phases. In CAMP-I, we concentrated on comprehending the intricate features of MR images, generating a fresh collection of synthetic MR images, inclusive of both MGMT methylated and MGMT unmethylated statuses. The subsequent phase CAMP-II involved the construction of sparsity limitations through the modeling of an adaptive sparse penalty. This was coupled with the incorporation of transfer learning to predict the methylation status of the MR slices introduced. The adaptive sparse penalty's behavior self-modifies along with the learning parameters, adapting to variations in the data, such as contrast changes and tumor location shifts in MR images. To validate our algorithm's efficacy, we compared it with the most popular MR image dataset currently available. All our experiments adhered strictly to recognized scientific standards and were executed within a meticulously planned experimental environment. In the subsequent sections, we delve deeper into the dataset, training guidelines, hyperparameters, and our primary observations.

### A. Dataset

Our chosen dataset was the widely recognized RSNA-MICCAI brain tumor radiogenomic dataset [21]. Our rationale for this choice was multifaceted first, its vastness and diversity assured the reliability and broad applicability of our conclusions. Second Given its widespread use in research, it guarantees trustworthy and impartial results. 3) Pertinently, for the MGMT methylation status challenge, the dataset offers clear-cut features, precise annotations, minimal disruptions, and critically, it aligns with ethical standards, especially concerning data privacy and consent.

This data set includes records of 585 patients with brain tumors. Every MRI image in this collection has a resolution of 256x256 pixels, translating to a whopping 65,536 unique pixels for each image. The dataset is equipped with four separate modalities: FLAIR, T1w, T2w, and T1wCE. Remarkably, for every patient, their MR image is accessible across all these imaging modalities.

The methylation status for each image is ascertained experimentally and is documented in the dataset. Furthermore, the dataset is bifurcated based on the genetic attributes of the tumors. Prior to any analysis, MRI images were subjected to preprocessing to harness pertinent data and discard extraneous information. Only those slices with the essential brain segment details vital for determining MGMT methylation status were preserved. These chosen slices underwent further enhancement through multiple preprocessing stages, were transitioned to grayscale, and subsequently fed into the model.

### B. Training Protocols and Hyperparameters

The CAMP model, a Convolutional Neural Network (CNN) depicted in the table, is intricately designed for feature learning. Initiating with an input layer tailored for 256x256 single-channel images, it unveils its architecture with convolutional layers for nuanced feature extraction, enriched with batch normalization for network stabilization and training acceleration. Ensuing layers include LeakyReLU activation functions introducing non-linearity, and MaxPooling for spatial down-sampling, optimizing computational efficiency. LeakyReLU activation functions are strategically interspersed within the architecture, imparting non-linearity and ensuring the network's resilience against vanishing gradient issues by allowing a small gradient when the unit is not active. Dropout layers intersperse the architecture, enhancing generalization by mitigating overfitting risks through randomized input nullifications during training. In the network's evolution, UpSampling layers emerge, elevating spatial resolutions and recapturing lost spatial intricacies. Culminating in an output layer, the network synthesizes and presents its learned features, reflecting an output congruent with the input's spatial dimensions, but imbued with enriched, learned features. The architectural complexity, manifested in its trainable parameters, predominantly resides in its convolutional layers, driving the network's robust feature-learning capacity.

### C. Performance and Evaluation Metrics

- RMSE (Root-Mean-Square-Error): RMSE (5) measures the pixel-wise differences between the actual image and the synthetic or reconstructed image generated by the CAMP-I. A lower RMSE value indicates that the model's predictions are closer to the actual values, meaning the synthetic or reconstructed image is more similar to the actual image.

$$\text{RMSE} = \sqrt{\frac{1}{M \times N} \sum_{i=1}^{M} \sum_{j=1}^{N} (O_{ij} - G_{ij})^2} \quad (5)$$

where, $O_{ij}$ represents the pixel intensity at location ($i, j$) in original image and $G_{ij}$ represents the pixel intensity at location in the generated image ($i, j$) and $MXN$ total number of pixels in each image.

- Accuracy (6) is a metric used here to evaluate the performance of CAMP-II, including those predicting MGTM (O6-methylguanine–DNA methyltransferase) status in medical diagnoses. When predicting MGTM status, a CAMP-II classifies each case as either having a methylated MGTM promoter or an unmethylated MGTM promoter based on input features.

$$\text{Accuracy} = \frac{TP + TN}{TP + TN + FP + FN} \quad (6)$$

It calculates the ratio of correctly predicted instances to the total instances.

- Sensitivity (7) measures the ability of a model to correctly identify the actual positive cases among all the positive cases it predicts. A positive case here refers to the presence of a methylated MGMT promoter. However, accuracy can be misleading in imbalanced datasets, where one class significantly outnumbers the other, and for that reason, sensitivity is also used

$$\text{Sensitivity} = \frac{TP}{TP + FN} \quad (7)$$

- Specificity (8) focuses on the model's ability to correctly identify negative cases, and in the context of MGMT status prediction, a negative case refers to an unmethylated MGMT promoter.

$$\text{Specificity} = \frac{TN}{TN + FP} \quad (8)$$

True Negative Rate gauges the proportion of actual negatives that are correctly identified. It's crucial in contexts where the cost of a false positive is high, such as in certain diagnostic tests where a false positive could lead to unnecessary treatments or procedures.

## IV. RESULTS

In the this section, we present a comprehensive evaluation of the proposed Convolutional Autoencoders for MGMT Methylation Status Prediction (CAMP) method, assessing its performance in predicting the MGMT methylation status from MRI images of glioblastoma patients. The outcomes of our study highlight the effectiveness of CAMP in addressing the challenges associated with predicting MGMT methylation status, particularly in the context of the intricate and variable patterns observed in magnetic resonance imaging (MRI). We delve into the performance metrics, including root mean square error, accuracy, specificity, and sensitivity, to provide a detailed assessment of the model's predictive capabilities. Furthermore, we compare the results obtained by CAMP with benchmarked datasets, demonstrating its superior performance with an impressive scores across all metrics. These findings underscore the promising potential of CAMP in advancing the interpretation of MRI data for predicting MGMT methylation status in the context of glioblastoma, offering valuable insights for personalized treatment strategies.

## A. Phase-I Loss Curves: Multimodal MRI Image Synthesis using the CAMP Model

In Phase-I of the CAMP model training, the validation loss for FLAIR MRI synthesis reached an impressive low of 0.008, as shown in Figure 3. This value signifies the model's adeptness in capturing the distinct features and contrasts inherent in FLAIR images. The gradual reduction in the loss reflects the learning progression, suggesting that the convolutional autoencoder successfully minimizes the root mean squared differences between the synthesized FLAIR images and the ground truth. The meticulous attention to detail in FLAIR modality synthesis is crucial, given its significance in detecting abnormalities related to glioblastoma, and the model's ability to achieve such a low loss underscores its proficiency in this specific imaging domain.

Simultaneously, the validation losses for T1w and T2w modalities are observed to be 0.006 each, showcasing a commendable consistency in the model's performance across different image types. The low loss values for both T1w and T2w suggest that the convolutional autoencoder effectively captures the nuanced contrasts and spatial structures characteristic of these modalities. This uniformity in performance across diverse imaging modalities is a promising indication of the model's versatility in handling various imaging challenges, laying a solid foundation for subsequent phases of the CAMP model.

## B. Overall Performance Evaluation

We outline the summarized results of CAMP model in Figure 4 for predicting the MGMT (O-6-methylguanine-DNA methyltransferase) promoter methylation status, which is a relevant biomarker in certain types of cancer, such as glioblastoma. The model was trained and tested on a dataset comprising 13,716 MRI images to predict the MGMT status. We used ten fold cross validation settings to run our experiments.

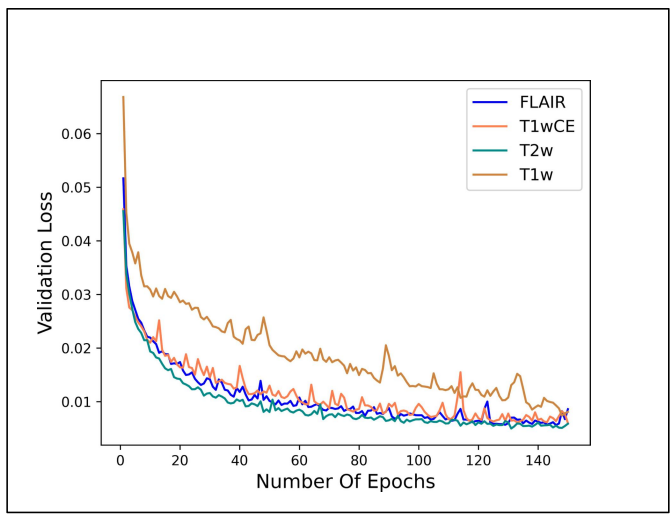

Fig. 3: Validation loss of MRI image synthesis during Phase-I of the CAMP model when assessed across four modalities (FLAIR, T1w, T2w, T1wCE) using root mean squared loss.

Overall the CAMP model achieved an accuracy of 97.8%. This means that, on average, the model correctly predicts the MGMT status in 97.8% of the cases in the testing set. The F1 score, a composite metric weighing precision and recall, stands at an impressive 97.9%. This signifies a robust balance between correctly identifying positive cases (precision) and capturing the majority of actual positive cases (recall). In the realm of medical diagnostics, achieving a high F1 score is vital, as it indicates a model's ability to strike an effective equilibrium between minimizing false positives and ensuring comprehensive identification of true positive cases.

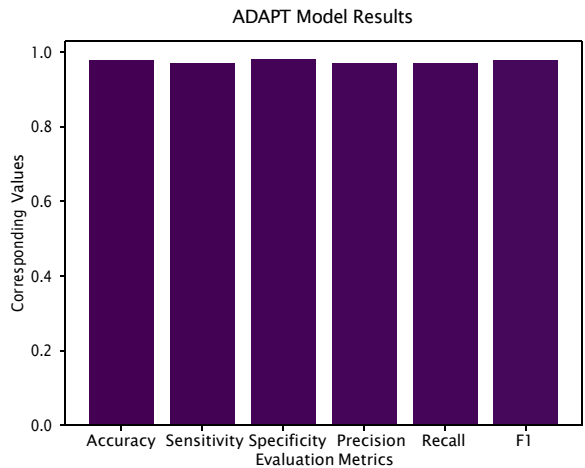

Fig. 4: A bar chart illustrating accuracy, precision, recall, sensitivity, specificity, and F1 score, achieved by the CAMP Model, showcasing the comprehensive evaluation of the model's effectiveness in predicting MGMT methylation status in Glioblastoma using MR scans.

Sensitivity, also known as recall or the true positive rate, attains a noteworthy value of 98%. This means that the CAMP model is effective in capturing instances of MGMT methylation, reducing the risk of false negatives. This is particularly important in medical applications, as missing cases of MGMT methylation could have significant clinical implications for the diagnosis and prognosis of brain tumors. This metric signifies the model's capability to accurately identify 98% of true MGMT-positive statuses, a critical aspect in medical diagnosis where identifying individuals with the condition is of paramount importance.

On the other hand, we observed a specificity value of 97%. This indicates that our model excelled in accurately identifying cases where MGMT was not methylated, offering a reliable tool for clinicians to rule out the presence of MGMT methylation in specific individuals. Specificity is a crucial metric in medical imaging research as it assesses the model's proficiency in distinguishing healthy individuals or cases without the condition of interest. Such robust specificity values enhance the clinical utility of our predictive model, contributing to the advancement of personalized treatment strategies for patients with glioblastoma.

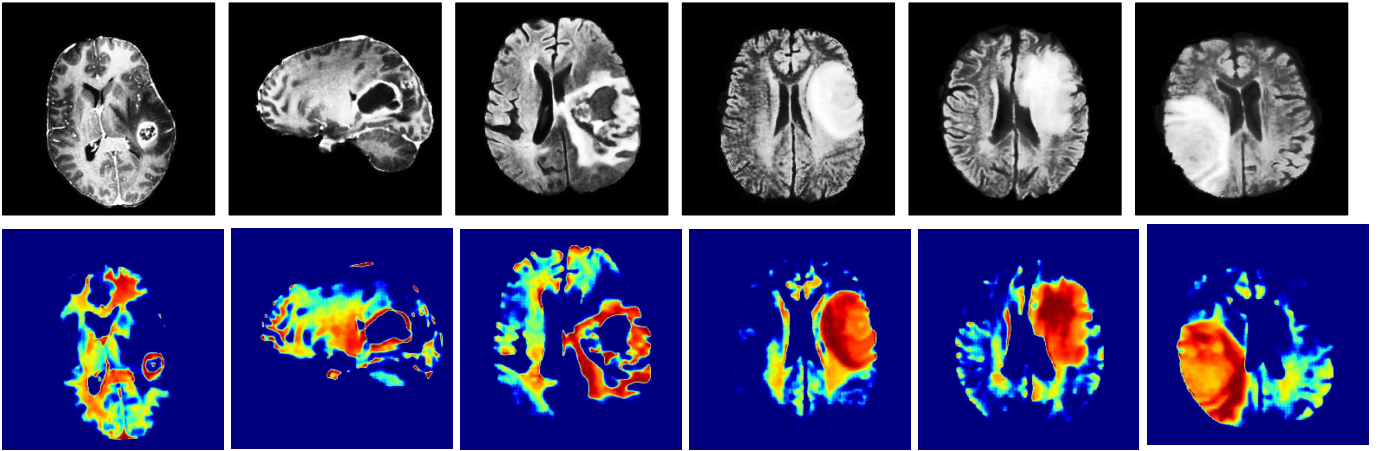

Fig. 5: (a) Row one shows some example pre-processed images of various MRI Modalities (T1w, T2w, FLAIR, T1wCE). (b) Row two shows the activation Maps of CAMP Model for Predicting MGMT Methylation Status in Glioblastoma. The maps provide visual insights into the convolutinoal neural network's feature activation patterns during the prediction of MGMT methylation status in Glioblastoma, contributing to a deeper understanding of the model's decision-making process.

## V. Discussion

The results of our study present a significant advancement in predicting the MGMT methylation status in glioblastoma patients through the application of Convolutional Autoencoders for MGMT Methylation Status Prediction (CAMP). The primary objective of our research was to address the challenges associated with the uneven contrast appearance, variability within lesions, and irregular enhancement patterns in MRI images connected with the MGMT signature.

The accuracy of our proposed method is noteworthy, achieving an accuracy score of 0.97, a specificity of 0.98, and a sensitivity of 0.97. These results indicate a high precision in predicting the MGMT methylation status, showcasing the robustness and effectiveness of the CAMP approach. The achievement of such accuracy is particularly crucial in the context of glioblastomas, considering their aggressive nature and the imperative need for personalized treatment plans.

Our study successfully demonstrates the capability of the autoencoder neural network in generating synthetic MRI slices that faithfully preserve the intricate structures of brain tissue, fat, and individual tumor structures across various MRI modalities. This not only enhances the interpretability of the MRI data but also addresses the challenges posed by the uneven contrast appearance and variability within lesions.

Figure 5(a) displays a set of example pre-processed images representing various MRI modalities, including T1-weighted (T1w), T2-weighted (T2w), Fluid-Attenuated Inversion Recovery (FLAIR), and T1-weighted post-contrast enhanced (T1wCE) images. These images constitute the input data to our Convolutional Autoencoders for MGMT Methylation Status Prediction (CAMP) model. Pre-processing is a critical step in ensuring the uniformity and compatibility of the data across different modalities, thereby enabling the model to effectively learn and extract relevant features from diverse MRI sequences. The inclusion of multiple modalities enhances the model's ability to capture the nuanced patterns associated with MGMT methylation status, contributing to the overall robustness and accuracy of the predictive capabilities.

In Figure 5(b), the second row showcases the activation maps generated by the CAMP model during the prediction of MGMT methylation status in glioblastoma. These activation maps provide visual insights into the feature activation patterns of the convolutional neural network (CNN) at different layers. By visualizing the areas of the input image that contribute most to the model's decision-making process, these maps offer a deeper understanding of the internal representations learned by the CNN. The distinctive patterns and spatial activations depicted in the maps shed light on the regions of interest and emphasize the model's focus on specific features relevant to MGMT methylation status prediction. This interpretability aspect is crucial for establishing trust in the model's predictions and can facilitate further exploration of the biological significance of the identified features, fostering transparency and confidence in the clinical application of the CAMP model.

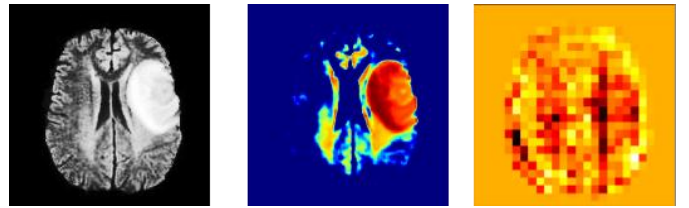

Fig. 6: An example MRI image, an activation map during image synthesis (CAMP-I phase), and activation maps of (CAMP-II phase) taken from the final layers and scaled. The tumor region is characterized by a prevalence of dark pixels in both activation maps, underscoring the critical importance of these darker pixels in the decision-making process.

### A. Effect of Adaptive Sparse Penalty

The incorporation of an adaptive sparse penalty in our model further contributes to the success of our approach. The adaptive nature of the sparse penalty allows for adjustments in response to variations in data, such as changes in contrast and tumor location in MR images. This adaptability is a critical factor in dealing with the complexities inherent in glioblastoma cases, contributing to the superior performance of our method in predicting MGMT methylation status.

The effect of adaptive sparse penalty can be seen in Figure 6 in which we show an example MRI image, a heatmap during image synthesis phase, and an activation map during MGMT status prediction phase. After incorporating sparse penalty, the activation map highlights areas of significance, with a particular focus on the regions associated with the tumor. The concentration of dark pixels in the activation map corresponds to the critical features extracted by the autoencoder, emphasizing the importance of these regions in influencing the synthetic MRI slices. This map serves as a visual representation of the model's attention, providing insights into the specific areas that contribute significantly to the accurate synthesis of MR images.

Likewise, the activations in the last layers of our model are instrumental in predicting the MGMT status. The dark pixels in this representation align with the tumor location, emphasizing their pivotal role in the decision-making process of the model. The discernible pattern of dark pixels in the last layer activations reinforces the model's ability to recognize and prioritize features associated with the MGMT status. This visualization not only enhances our understanding of the model's internal workings but also underscores the significance of dark pixel prevalence in informing accurate predictions of MGMT methylation status in glioblastoma patients' MRI images.

*B. Comprarative Analysis*

We conducted a thorough analysis of our approach in relation to contemporary computational techniques designed for predicting MGMT methylation status using MRI images. Specifically, we chose to benchmark our methodology against the recent work of Thi et al. outlined in [22]. The rationale behind selecting this study lies in its state-of-the-art machine learning (ML) model grounded in radiomics to predict MGMT methylation status. Thi et al. demonstrated innovation by creating twenty-five novel features, extracting information from multimodal MRI images of GBM patients, and incorporating annotated MGMT methylation status into their model. This choice for comparison aligns with our goal of assessing the effectiveness of our approach in the context of advancements in the field, allowing for a comprehensive evaluation of its performance against cutting-edge methodologies.

Thi's method, as referenced, achieved an Area Under the Curve (AUC) of 0.93, indicating a commendable predictive capability as shown in Figure 7. Our study, however, surpasses this benchmark with a notable performance improvement, achieving an AUC score of 0.97. This substantial increase in prediction power is attributed to the adaptive sparse penalty that penalizes activations that are focused on areas of the image that don't contribute to the MGMT methylation status. The high accuracy underscores the efficacy of the Convolutional Autoencoders for MGMT Methylation Status Prediction (CAMP) in enhancing the precision of predicting MGMT methylation status. The comparative evaluation extends beyond AUC, considering other metrics as well. Notably, our approach demonstrated an accuracy score of 0.97, showcasing the robustness and reliability of our model in classifying MGMT methylation status. Additionally, the specificity of our approach stands at 0.98, emphasizing its ability to correctly identify cases where MGMT is methylated. The sensitivity of 0.97 further highlights the model's proficiency in detecting cases where MGMT is not methylated.

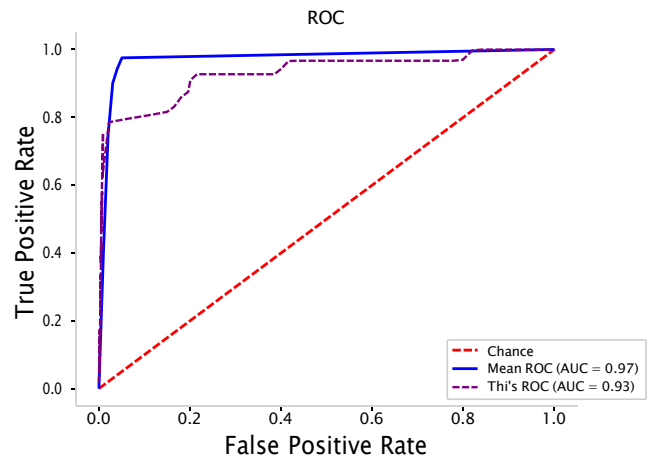

Fig. 7: Illustration of the Receiver Operating Characteristic (ROC) curve comparing the performance of the proposed CAMP approach (AUC = 0.97) with the *Thi's* approach (AUC = 0.93) for predicting MGMT methylation status. The higher area under the curve (AUC) indicates the superior discriminatory ability of our approach in comparison to the *Thi's* approach [22].

These results collectively position our proposed method as a highly promising and superior approach compared to Thi's, providing a substantial advancement in the interpretation of MRI data for predicting MGMT methylation status in brain cancer patients. The elevated AUC and other performance metrics underscore the potential clinical significance and reliability of the Convolutional Autoencoders for MGMT Methylation Status Prediction (CAMP) in the context of glioblastoma management.

The outcomes of our research hold significant implications for clinical practice. The high accuracy and specificity of our method make it a promising tool for clinicians in assessing the MGMT methylation status of glioblastoma patients. This, in turn, can inform and guide the development of tailored treatment plans, optimizing therapeutic outcomes. In conclusion, our study introduces a novel and effective method, CAMP, for predicting MGMT methylation status in glioblastoma patients based on MRI images. The robustness, adaptability, and high accuracy of our approach underscore its potential as a valuable tool in the clinical setting, contributing to improved patient prognosis and personalized treatment strategies.

## VI. Conclusion

In this work, we introduced a Convolutional Autoencoders for MGMT Methylation Status Prediction (CAMP) model, to address the challenges of predicting MGMT methylation status in glioblastoma patients using magnetic resonance imaging (MRI). Leveraging deep learning, our approach generates synthetic MRI slices with a tailored sparse convlutional autoencoder, incorporating an adaptive sparse penalty to enhance predictive accuracy. The method exhibits substantial improvements in MRI image synthesis, preserving tissue structures across modalities. Achieving an accuracy score of 0.97, specificity of 0.98, and sensitivity of 0.97, CAMP outperforms benchmarked datasets, demonstrating its potential to enhance the interpretation of MRI data for personalized treatment planning in glioblastoma patients. In conclusion, our findings underscore the significance of CAMP in providing a more accurate and tailored method for assessing the critical MGMT methylation status. This contribution aligns with the broader goal of customizing treatment plans and improving prognostic outcomes for glioblastoma patients. Future research may explore the integration of CAMP into clinical practice, advancing the prospects of personalized medicine in the context of glioblastoma management.